%% file: 0_main.tex
\documentclass[conference, 10pt, letterpaper]{IEEEtran}
\IEEEoverridecommandlockouts

\usepackage{graphicx}
\usepackage{xspace}
\usepackage{subcaption}
\usepackage{optidef}
\usepackage{booktabs}

\newcommand{\system}{\texttt{NEO-GNN}\xspace}

\author{
    Aruna Jayarajan, N. Cameron Matson, and Karthikeyan Sundaresan\\
    School of Electrical and Computer Engineering, Georgia Institute of Technology, USA\\
    \texttt{\{ajayarajan6, ncmatson\}@gatech.edu, karthik@ece.gatech.edu}
}

\begin{document}
\title{LEO Satellite Network Orchestration with Heterogeneous Graph Neural Networks
}

\maketitle

\input{01_abstract}
\input{1_introduction}

\input{2_motivation}

\input{3_approach}

\input{4_evaluation}

\input{5_related_work}

\input{6_conclusion}

\section*{Acknowledgment}
{
This work was supported in part by NSF (CNS 2208761).
}

\clearpage

\bibliographystyle{IEEEtran}
\bibliography{neognn}

\end{document}

%% file: 01_abstract.tex
\begin{abstract}
Low Earth Orbit (LEO) satellite constellations are becoming essential for expanding global Internet access, especially in remote and under-served areas. 
However, their highly dynamic nature, arising from network mobility, introduces complex coordination challenges between the dynamic satellites and the ground nodes (gateways and terrestrial devices).
This is underscored by limited satellite visibility windows and spatially imbalanced user traffic demands. 
Local association (cell-satellite-gateway) strategies, such as nearest-satellite or greedy load-based selection, result in partial terrestrial coverage or lead to load imbalance that affects traffic demand fulfillment. Network-driven orchestration through centralized optimization can strike an efficient balance between these two key objectives, but is often computationally intensive for periodic operation and real-time deployment. This work presents a learning-based network orchestration framework, \system, that models a satellite-ground network as a dynamic spatio-temporal graph. In contrast to prior works, it employs a heterogeneous Graph Neural Network (GNN), where satellites, gateways, and ground cells are modeled as distinct node types to capture their varied visibility and networking capabilities. They are trained in an unsupervised manner using tailored loss functions to balance the dual requirements of coverage and utilization, and produce efficient, real-time association decisions during inference.
Evaluations show that \system delivers complete ground-cell coverage, improves traffic demand satisfaction through balanced satellite and gateway use, and remains robust under dynamic visibility and partial satellite failures. \system provides a scalable and efficient alternative to traditional optimization methods for real-time network orchestration in bent-pipe LEO satellite systems.
\end{abstract}

%% file: 1_introduction.tex
\section{Introduction}

Low Earth Orbit (LEO) satellite constellations and networks (LSNs) are becoming an integral part of the global internet infrastructure. They consist of thousands of fast-moving satellites designed to deliver broadband connectivity, especially in regions where terrestrial networks are sparse or impractical to deploy. 
Many current constellations adopt a \textit{bent-pipe} or \textit{single-hop} architecture, where terrestrial user traffic is relayed via satellites to ground-based gateways (GW)  without inter-satellite link (ISL) data forwarding. While some modern systems are beginning to incorporate onboard processing capabilities to enable ISLs, bent-pipe designs remain attractive due to their lower latency and hardware simplicity~\cite{liu-leosatelliteconstellations-2021}.

\begin{figure}[t]
    \centering
    \includegraphics[width=\linewidth]{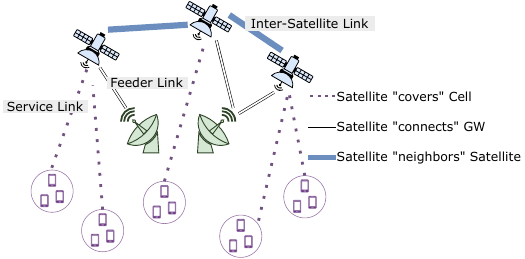}
    \caption{Heterogeneous Graph for GNN-based Orchestration.}
    \label{fig:graph_schema}
\end{figure}

Regardless, LSN performance is largely bottlenecked by the capacity of their service (cell-satellite) and feeder (satellite-gateway) links (Fig.~\ref{fig:graph_schema}). Hence, to achieve maximum network efficiency of an LSN, it becomes critical to balance the traffic demand carried across satellites and GWs to maximize network capacity/utilization, while simultaneously ensuring coverage of all the terrestrial cells. This challenging dual-objective requires the LSN system to carefully assign satellites to cover ground cells as well as route traffic through available GWs in a dynamic manner that keeps pace with the satellite mobility, while accounting for varying satellite visibility, user traffic demand, and gateway/satellite capacity limits. We refer to this problem as the network orchestration problem in LSNs. 

Simple assignment (association) strategies that rely only on local information, such as assigning cells to the nearest satellite or satellites to the least-loaded GW, are fast and easy to implement. 
However, not having a network-wide perspective often results in poor load distribution and network utilization or incomplete cell coverage by such schemes.
On the other hand, advanced solutions that employ global optimization can provide significantly enhanced performance by accounting for the entire network, but are computationally expensive to run in real time.  We show it takes over a minute to solve the optimization for a 4750 node network, which is too slow to serve as a centralized controller that must compute frequent orchestration updates on the order of seconds.
Thus, we desire an approach that can bring the superior performance of network-driven strategies, but can also scale and operate in real-time for practical deployment in dynamic LSNs. 

To this end, we propose a heterogeneous Graph Neural Network (GNN)–based network orchestration framework, called \system. It operates in real-time to determine configurations (assignments) that strike an efficient balance between satisfying traffic demands (network utilization) and cell coverage even in the face of satellite dynamics. The intuition behind \system is that LEO satellite orbital trajectories exhibit spatio-temporal structures that can be learned over time by training on network snapshots that capture the evolving satellite-ground topology. This can help understand how individual satellites should behave in terms of their associations (to cells and GWs) based on where they are so as to balance both utilization and coverage, while respecting visibility and capacity constraints. This work focuses on showing how role-aware graph learning can be applied in practice to large-scale bent-pipe LEO constellations under realistic mobility and capacity constraints.
While conventional GNNs for LSNs~\cite{chen-deepreinforcementlearningbased-2025} consider a homogeneous set of nodes, and are unable to cater to both utilization and coverage, \system  models the LSN as a dynamic, ``heterogeneous'' graph (like the one shown in Fig.~\ref{fig:graph_schema}) where satellites, ground cells, and GWs are modeled as distinct node types to capture their varied visibility, networking capabilities and constraints, thereby enabling more accurate and flexible decision-making. In addition, \system's heterogeneous GNN is trained using loss functions that are tailored to balance the dual requirements of coverage and utilization. By learning satellite association behavior offline, \system enables scalable and resilient real-time decisions.

\system is evaluated on realistic constellations (similar to those used by Starlink).  
The results demonstrate that \system significantly outperforms both local and network-driven baseline approaches (20–30\% more demand satisfaction, 15–25\% more coverage) and closely approximates the performance of global optimization-based solutions, in catering to both utilization and coverage. It delivers real-time inference (under 620 ms for a 4750-node network) that scales to larger constellations, unlike optimization approaches, making it well-suited for deployment in dynamic LSN systems. Its unsupervised nature brings resilience, allowing it to sustain performance even during appreciable changes in satellite constellation.

%% file: 2_motivation.tex
\section{Motivation}
\label{sec:motivation}
In terrestrial networks, there are two types network nodes: mobile users and static base stations.  The typical cellular deployment make user association and mobility induced handovers simple: users typically are connected to the closest BS and handed over when they move closer to another BS. 

LSNs present a fundamentally different paradigm.  Rather than two types of nodes, there are three: the user, the satellite, and the satellite GW.  

Because of their altitude, the ground coverage area of satellites is much larger than terrestrial cells (even for the relatively low altitude LEO satellites).  Even with state-of-the-art high-throughput narrow spot beams~\cite{kim-downlinkanalysisevaluation-2024}, the coverage area of several satellites is likely to overlap on Earth.  This raises the question: If multiple satellites are visible from a user's vantage point on earth, and multiple GWs are visible from a satellite's location in space, how does the network operator decide the ``best'' association of user-satellite-GW?
Compounding this problem is the fact that satellites, which now represent a key infrastructure node, are highly dynamic, much more mobile than the terrestrial users.
The orchestration problem from a network perspective can be expressed as a two-sided assignment: which ground cells (and all its users) each satellite should serve (service side), and which gateways each satellite should use for backhaul (feeder side). 
These decisions must be made frequently as satellites move, necessitating an efficient, real-time decision process.

We identify two primary objectives desirable by a network orchestration scheme:
1) Full ground cell coverage, and
2) balanced utilization across the constellation to maximize network capacity.
These two objectives must be considered in the face of certain constraints: each satellite (service link) and GW (feeder link) has strict capacity limits that affect the total demand they can serve, and assignments must be made quickly or they risk becoming invalid (e.g. a satellite is no longer visible) due to satellite mobility.
Given the likely scenario that total user demand is greater than the capacity of the system, we can either choose to only serve a fraction of users or to serve a fraction of the demand for all users.
We now investigate two simple satellite assignment methods (local heuristics and global optimization) and demonstrate how they do a poor job in balancing the objectives given our constraints.

\begin{figure}[t]
    \centering
    \begin{subfigure}[t]{0.48\columnwidth}
        \centering
        \includegraphics[width=\linewidth]{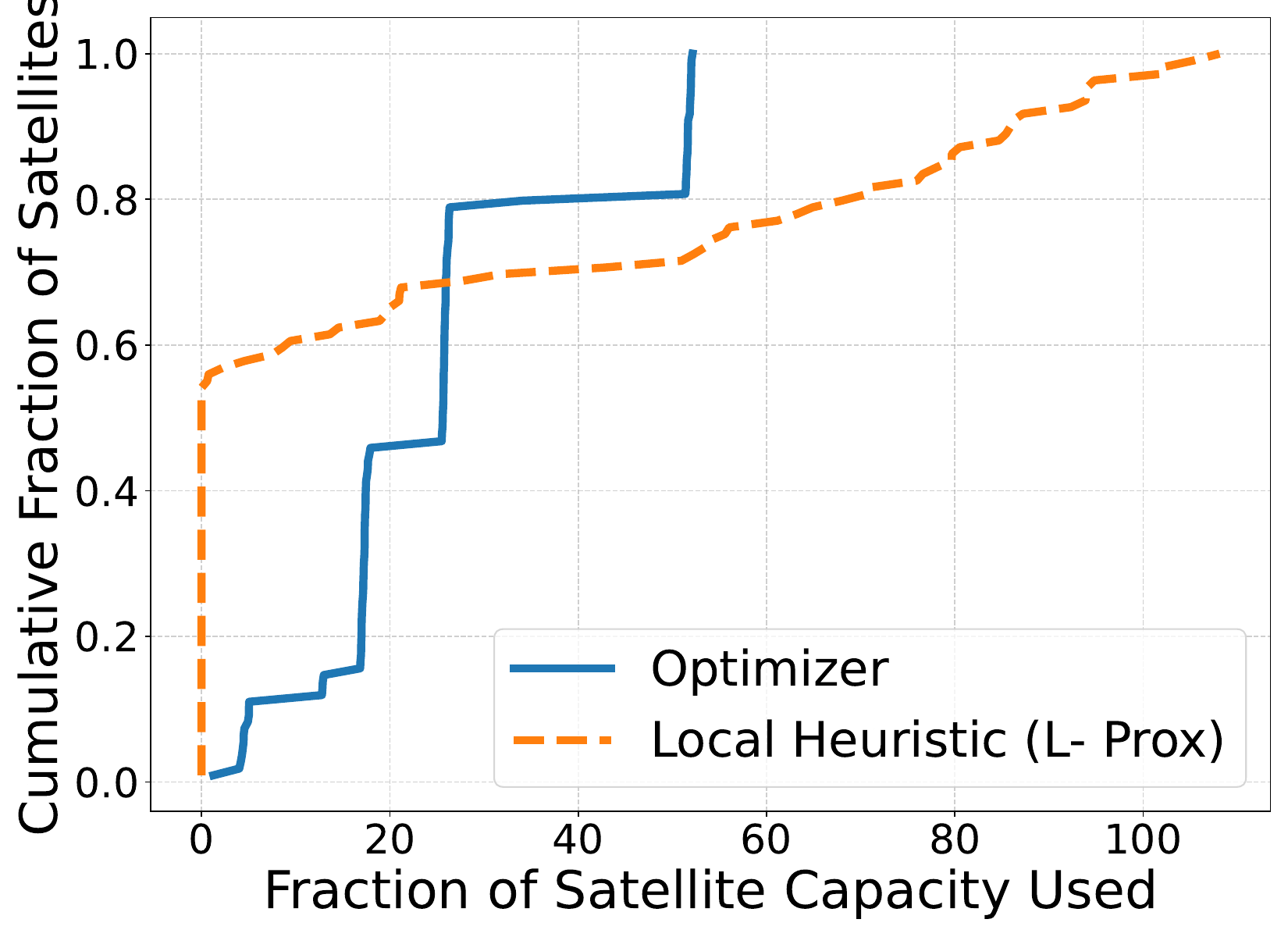}
        \caption{CDF}
        \label{fig:cdf_demand_heuristic}
    \end{subfigure}
    \hfill
    \begin{subfigure}[t]{0.48\columnwidth}
        \centering
        \includegraphics[width=\linewidth]{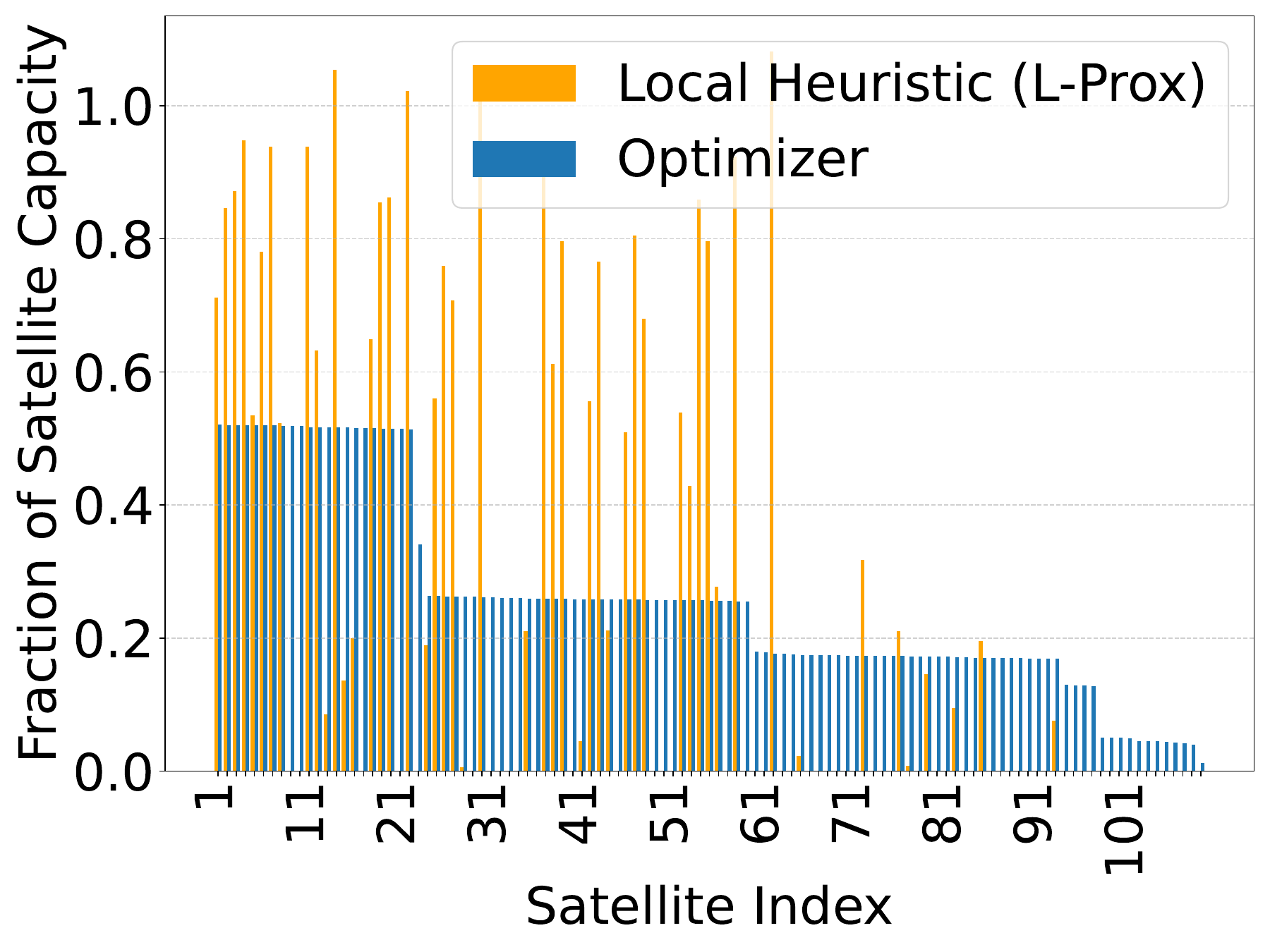}
        \caption{Demand plot}
        \label{fig:cdf_demand_optimizer}
    \end{subfigure}
    \caption{Comparison of local heuristic and optimizer.}
    \label{fig:cdf_demand_comparison}
\end{figure}

\textbf{Limitations of Local Heuristics}
Local assignment methods, such as cells/satellites selecting the nearest satellite or least-loaded GW respectively, are fast and easy to implement. However, they lack awareness of the broader view of the network state and make decisions independently, without accounting for overall system constraints. As a result, this setup frequently causes a few visible satellites or GWs to become overloaded, while others are left idle.

To demonstrate, we have conducted an experiment in which cell-satellite and satellite-GW association is made based exclusively on distance (Details in Section~\ref{sec:evaluation}).
The simulation is based on a constellation comprising 1,584 LEO satellites (similar to the Starlink Phase 1 deployment)\cite{fcc-applicationfixedsatellite-2020}, 54 ground GWs and 4,569 uniformly distributed user cells throughout the continental U.S. 
In this baseline model, each ground cell is assigned to the geographically closest satellite. Then,  each satellite is paired with its nearest GW to route data to the internet. This heuristic only considers the physical distance (and visibility), not the demand, congestion, or capacity constraints.
As illustrated in Figure~\ref{fig:cdf_demand_comparison}, such local heuristics lead to highly skewed demand distributions, with some satellites exceeding their capacity and many others left idle, showing significant load imbalance.
Even though full coverage is maintained, fewer than half of the satellites are actively serving cells, leading to serious under utilization of the available system capacity.

The problem with this load imbalance is that satellites and GWs have limited capacity, both in terms of physical resources allocated to them~\cite{liu-leosatelliteconstellations-2021} and their on-board processing~\cite{razmi-onboardfederatedlearning-2021}.  In the event that the demand assigned to a particular satellite exceeds this capacity, user throughput served can only be a fraction of their requested demand.
To illustrate, let the capacity of a satellite be $C$.  Each cell (defined by the set $\mathcal{U}$, indexed by $i$, $|\mathcal{U}| = N$) generates a demand $r_i$.  Through this local assignment some subset of cells $\mathcal{V} \subset \mathcal{U}$ is assigned to a particular satellite $j$.  The total demand assigned to that satellite is $\tilde R_j = \sum_{i \in \mathcal{V}} r_i$.  If $\tilde R_j > C_j$ the demand served for each cell must be throttled.  For fairness, assume each cell is allowed to transmit a fixed ``satisfaction'' ratio of its desired demand $\lambda \equiv C/\tilde R_j$.  Obviously, the more demand assigned to the satellite, the lower the satisfaction ratio.  Depending on the system, the satisfaction ratio may be the same or unique across satellites and gateways. Regardless, a load balanced system is needed to provide a higher demand satisfaction overall.

Alternatively, rather than allow each user to connect to their closest satellite, the satellites themselves could choose (or be controlled to choose) only as many users at a time not to exceed their capacity.  This way the served users would receive their full demand, but many users may receive no resources at all, which is not desirable either.

\textbf{Why Global Optimization Is Impractical}
Global optimization methods produce better-balanced solutions by solving large-scale constrained problems. 
However, they require complete, real-time knowledge of the network state and are computationally expensive.
Imagine instead of local decisions, a central controller load balances the entire network, and then broadcasts the results throughout the network.
We use the following formulation used for load balancing:
    {\small
    \begin{mini*}[]
        {x}{L}{}{}
        \addConstraint{\sum_s \sum_j x_{isj}(t) r_{j}(t)}{\le \eta L}{\forall i}
        \addConstraint{\sum_i \sum_j x_{isj}(t) r_{j}(t)}{\le L}{\forall s}
        \addConstraint{\sum_i \sum_k x_{isj}(t)}            {= 1}           {\forall j}
        \addConstraint{x_{isj}}                                    {\in \{0, 1\}}  {\forall i,s,j}
    \end{mini*}}
where $x_{isj}$ is the binary assignment variable for each flow from user $j$ through satellite $s$ to GW $i$.  The variable $\eta$ is used to adjust the difference in capacity between satellites and GWs.  The objective is to minimize the RHS variable $L$ in the inequality constraint.  This is equivalent to minimizing the maximum load on any satellite or GW.

We solve this multidimensional assignment problem\footnote{The multidimensional assignment problem is NP-Hard~\cite{pisinger-wherearehard-2005} thus, and furthermore there exist no known polynomial-time approximation algorithms.} to perform load balancing across the network using the commercial solver Gurobi~\cite{gurobioptimizationllc-gurobioptimizerreference-2026}.
Unsurprisingly, we see in Figure~\ref{fig:cdf_demand_comparison} that the median satellite load is much lower after optimization.
However, for a smaller problem than the one considered in the results of Figure~\ref{fig:cdf_demand_comparison} with only 600 users and 20 GWs, Gurobi takes 609~s to bound the solution to within 2\% of the optimal value.
Considering that LEO satellites are visible for a few minutes at most, it is infeasible to run this type of end-to-end, global optimization at the speed of satellite mobility, not to mention the small network size.
In Table~\ref{tab:quasi-global-runtime} we report the solve time of a ``quasi-global'' optimization on several network scenarios with varying numbers of GWs, cells, and satellites.  The quasi global optimizer solves the multidimensional assignment problem in two stages, first optimizing the cell-satellite connection, followed by satellite to GW.  Even solving this sub-optimal version of the problem for small networks takes a prohibitive amount of time, making such approaches  impractical for real-time deployment.

\begin{table}[t]
    \centering
    \caption{Quasi-Global Optimization Run Time}
    \begin{tabular}{c c c c c}
        \toprule
        \textbf{GW}  & \textbf{Cells} & \textbf{Satellites} &\textbf{Total Nodes}& \textbf{Opt. Solve Time} (s) \\
        \midrule
        15  & 1147  & 54    & 1216  & 2     \\
        21  & 1511  & 76    & 1608  & 10    \\
        54  & 4569  & 127   & 4750  & 64    \\
        \bottomrule
    \end{tabular}
    \label{tab:quasi-global-runtime}
\end{table}

\textbf{Why Learning-Based Models, and GNNs Specifically, Are Suitable}
LSNs are dynamic, structured, and partially observable. As satellites move along predictable orbital paths, visibility relationships between ground cells, satellites, and GWs evolve over time.  Since no single node has full network visibility, assignment decisions must rely on local context and real-time constraints.
Learning-based orchestration offers a promising alternative. By training on past visibility and demand patterns, a model can learn to make fast, effective decisions using local and structural features.
Graph Neural Networks (GNNs) are particularly well-suited for this task. They work directly on graph-structured data and support localized message passing, allowing satellites to coordinate based on their current neighborhood. A heterogeneous GNN can be employed to represent the distinct roles of satellites, GWs, and ground cells, enabling type-specific message passing and constraint-aware decision-making. The proposed model, \system, is built on this architecture and is described in the following section.

%% file: 3_approach.tex
\section{Design of \system}
\label{sec:approach}

\subsection{Overview}
\system represents the network as a heterogeneous graph, where each node type---satellite, GW, ground cell---has unique attributes and responsibilities. It uses a Heterogeneous Graph Neural Network (HetGNN) architecture~\cite{shi-heterogeneousgraphneural-2022}, with type-specific message passing and role-aware aggregation, to reason about connectivity and resource allocation. This design enables the model to generalize orchestration strategies not just from a single network snapshot, but across a sequence of dynamic graphs that reflect evolving constellations.

The model is trained without labeled assignments, using constraint-based losses to enforce coverage, capacity, and clear assignments. While described as unsupervised, the training is fundamentally driven by system feasibility, with penalties for uncovered cells, overloaded satellites, and ambiguous decisions.
This is done for two reasons: 
First, generating a large dataset of optimized labels is impractical given the scale of the problem. Second, the constraint-based formulation allows the system to adapt gracefully under stress, such as gateway congestion or satellite failures, while respecting physical and architectural limits.

The model is trained across many orbital snapshots. This ensures that embeddings capture temporal patterns, such as repeated passes and coverage gaps—and enables generalization to unseen scenarios. The result is a scalable, real-time orchestration approach that improves load balancing, gateway utilization, and coverage fairness.

\system is designed to quickly approximate the globally optimal cell–satellite–gateway assignment (over 100$\times$ faster than direct optimization), while effectively balancing the coverage–utilization trade-off that challenges local heuristics.

In the following sections we
first give a formal definition of the problem and underlying graph structure (Section~\ref{sec:problem_def}),
then we provide details of the GNN model at the heart of \system (Section~\ref{sec:gnn_arch}),
describe the multi-objective loss function used and the training methodology (Section~\ref{sec:loss}),
and finally we discuss deployment considerations (Section~\ref{sec:deployment}).

\subsection{Problem and Graph Definition}
\label{sec:problem_def}
Efficient orchestration in Low Earth Orbit (LEO) satellite networks presents a unique challenge due to the dynamic topology of the system, mobility of the infrastructure, and strict architectural constraints. 
As satellites move along their orbital trajectories, the visibility to ground cells and gateways changes frequently, complicating the task of maintaining continuous user coverage and achieving effective load distribution and traffic demand satisfaction.
The core problem is to assign each satellite a feasible subset of ground cells to serve and an appropriate gateway (or another satellite) for data egress. These assignments must satisfy several real-world constraints:

\noindent$\bullet$ 
\textbf{Visibility constraints:}
Satellite visibility is determined by the elevation angle of the satellite as observed from the ground. 
A satellite is considered visible only when its position is sufficiently above the horizon, exceeding a minimum elevation threshold that ensures reliable line-of-sight communication.
   
\noindent$\bullet$ \textbf{Capacity limits:} 
Satellites and GWs are subject to finite throughput limits, restricting the amount of traffic they can handle.

 \noindent$\bullet$ \textbf{Architectural constraints:}  
 Satellite topologies may be either bent-pipe or support ISLs which can 
 increase the flexibility in assignment.

As defined in Sec.~\ref{sec:motivation} each cell has a total desired traffic rate/demand, each satellite and GW has a corresponding capacity.  In the event that the total demand assigned to a satellite or GW exceeds that capacity, all of the users assigned to that node experience a drop in throughput  proportional to the overload.

Though satellites are always moving, we take the common approach (and backed by observed Starlink behavior \cite{tanveer-makingsenseconstellations-2023} and discretize time into small snapshots, in this case 20s intervals.
At each time index \( t \), the satellite network is represented as a typed graph:
\[
\mathcal{G}_t = (\mathcal{V}_t, \mathcal{E}_t), \quad
\mathcal{V}_t = \mathcal{S}_t \cup \mathcal{C} \cup \mathcal{B}, \quad
\mathcal{E}_t = \mathcal{E}_t^{\text{vis}} \cup \mathcal{E}_t^{\text{ISL}} 
\]
Where $\mathcal{S}_t$, $\mathcal{C}$ and $\mathcal{B}$ are the sets of satellite nodes (dynamic), cells, and GWs respectively.
$\mathcal{E}_t^{\text{ISL}}$ denotes logical satellite–satellite edges used only for message passing; our evaluation focuses on bent-pipe architectures without ISL data forwarding, while the edge subset \( \mathcal{E}^{\text{GSL}}_t \) is visibility-based edges connecting satellites to cells and GWs,
A GSL edge exists at time index $t$ (i.e. $e^{ij}_t \in \mathcal{E}^{\text{GSL}}_t$) if satellite $j$ is visible from ground node $i$ in the interval $[t, t+1)$.

\subsection{GNN Architecture}
\label{sec:gnn_arch}

\begin{figure}[t]
    \centering
    \includegraphics[width=\linewidth]{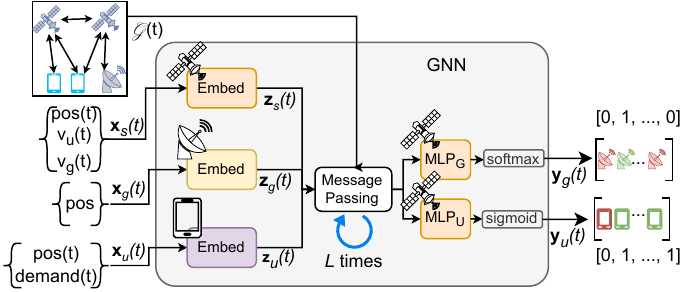}
    \caption{Forward Pass Through \system GNN}
    \label{fig:gnn_foward}
\end{figure}

The general structure of \system's GNN model is shown in Figure~\ref{fig:gnn_foward}.
Graph Neural Networks (GNNs) have been widely applied to satellite problems such as routing~\cite{gao-topologycompresseddatadelivery-2025, liu-routingsmallsatellite-2021}, scheduling~\cite{casadesus-vila-autonomouscooperationheterogeneous-2024}, and handover management~\cite{lee-handoverstrategyleo-2025}. However, these models generally rely on \textit{homogeneous} graph formulations, where all nodes and edges are treated the same. This overlooks the distinct roles and constraints of each network entity, where ground cells act as passive demand sources, gateways provide fixed-capacity backhaul, and satellites operate as mobile relays making assignment decisions under visibility constraints. Such a distinction becomes critical, when catering to the dual objective of coverage and utilization. 

In contrast, Heterogeneous GNNs are designed to model graphs with multiple node and edge types, allowing role-specific processing and type-aware message passing.
Although they have been explored in satellite resource scheduling~\cite{wang-learningmultisatellitescheduling-2024}, their use for real-time orchestration under dynamic visibility, mobility, and architectural constraints remains limited.
The model presented in this work uses:
typed \textbf{nodes} with separate embeddings for satellites, gateways, and ground cells.
typed \textbf{edges} with separate message functions for satellite–cell, satellite–gateway, and satellite–satellite interactions, and
\textbf{role-aware aggregation} allowing satellites to aggregate messages differently depending on the neighbor type.

The GNN works as follows:
Given the graph $\mathcal{G}_t$ for a specific snapshot of the network, for each node \( v \in \mathcal{V}_t \) we have a time-dependent feature vector \( \mathbf{x}_v(t) \).
The contents of each feature vector depends on the specific role:
For cells the feature vector, $\mathbf{x}_u$, is position and its time varying demand;
for GWs, $\mathbf{x}_g$ is position.
and for satellites the feature vector, $\mathbf{x}_s$, include the time-varying position as well as multi-hot encoded vectors indicating the visibility to cells and GWs.
These raw feature vectors are fed into node-specific learnable \textbf{embedding}.
Each embedding projects the raw feature vectors into a 64~dimensional latent space as $\mathbf{z}_u$, $\mathbf{z}_g$, and $\mathbf{z}_s$, respectively.

The next phase of the forward pass through the GNN is ``message-passing'' described in Eq~\ref{eq:msg_passing}.
\begin{align}
    \mathbf{z}_{s,i}^{(l+1)} = \sigma\Big(&
    W_{\text{self}} \mathbf{z}_{s,i}^{(l)} +
    W_{\text{cell}} \sum_{k \in \mathcal{N}_{\text{cell}}(i)} \mathbf{z}_{u,i} \nonumber \\
    &+ W_{\text{gw}} \sum_{m \in \mathcal{N}_{\text{gw}}(i)} \mathbf{z}_{g,m} +
    W_{\text{sat}} \sum_{j \in \mathcal{N}_{\text{sat}}(i)} \mathbf{z}_{s,j}^{(l)}
    \Big) \label{eq:msg_passing}
\end{align}
Message passing is equivalent to the convolution operation in CNNs in that its an aggregation of information among ``nearby'' nodes; however, whereas in CNNs the notion of nearby is restricted to fixed filter geometries, in GNNs the neighborhood is defined by the graph structure.
The message passing procedure is repeated multiple times which allows information to flow from multi-hop neighbors to each node.
In each message passing round, the embedding of a satellite $i$, $\mathbf{z}_{s,i}$ is updated by aggregating the embeddings of itself, its neighboring satellites ($\mathbf{z}_{s,j} \quad \forall j, e_{i,j} \in \mathcal{E}_t^{\text{ISL}}$), and visible cells and GWs.  Each of these is then multiplied by a unique, learnable, square (to preserve the dimension) weight matrix $W$.
The sum of these projections is then passed through a non-linear activation function $\sigma$ (e.g. ReLU).

After $L$, rounds of message passing the final embeddings ($z_s^{(L)}$) of each satellite node are fed in parallel multi-layer perceptrons (MLPs) and non-linear activation function to obtain the estimated cell coverage list and target node (either GW or another satellite if no GWs are visible).
The output shape of $\text{MLP}_U$ is equal to the number of cells, and the non-linear activation is sigmoid, i.e. the output vector for each satellite node $\mathbf{y}_g$ is a list of soft probabilities that the satellite will serve each cell.
Similarly, the output shape of $\text{MLP}_G$ is equal to the combined total of the number of GWs and available feeder satellites (i.e. the number of satellites which are GW visible).  Since each satellite may only connect to one target, off-loading node, the activation is softmax.  Note that in the event of a bent-pipe deployment, we simply restrict the output to the number of GWs.

\subsection{Constraint-based Loss and Training}

\label{sec:loss}

\subsubsection{Losses}
The training objective is built to reflect real-world constraints directly within the learning process. Rather than relying on separate optimization after inference, the model learns to make practical and efficient assignment decisions by minimizing a weighted combination of several loss components. The losses fall into three main categories: (i) \textit{Coverage Assurance Losses}, which ensure that all ground cells receive service and all gateways are actively used; (ii) \textit{Capacity Constraint Losses}, which prevent any satellite or gateway from being overloaded; and (iii) \textit{Unique Assignment Losses}, which push the model toward feasible decisions, like choosing just one gateway per satellite or assigning each ground cell to a single satellite.

Each part of the loss function reflects a specific operational need, and together they help the model make assignment decisions that are practical, efficient, and well-balanced. The following notation is used:

\begin{table}[h]
\centering
\caption{Summary of Notations}
\begin{tabular}{ll}
\toprule
\textbf{Symbol} & \textbf{Meaning} \\
\midrule
$N_s$ & Number of satellites \\
$N_c$ & Number of ground cells \\
$N_g$ & Number of gateways \\
$p^{\text{cell}}_{ij}$ & Probability satellite $i$ serves cell $j$ \\
$v^{\text{cell}}_{ij}$ & Visibility: satellite $i$ to cell $j$ (binary) \\
$p^{\text{gw}}_{ik}$ & Probability satellite $i$ uses gateway $k$ \\
$v^{\text{gw}}_{ik}$ & Visibility: satellite $i$ to gateway $k$ (binary) \\
$d_j$ & Traffic demand for cell $j$ \\
$C_{\text{max}}$ & Maximum satellite capacity \\
$G_{\text{max}}$ & Maximum gateway backhaul capacity \\
\bottomrule
\end{tabular}
\label{tab:notations}
\end{table}

\textbf{Coverage Losses}
These terms ensure all key entities in the network, ground cells and gateways, are actively served.

The coverage loss for cells is given in Eq.~\ref{eq:cell-coverage-loss}.
A similar loss $\mathcal{L}_\text{gw-cov}$ exists for GWs as well.
This loss encourages \textit{at least one} satellite to cover each cell.  The individual loss for each cell will be 0 if the total probability over all visible satellites is at least 1.
\begin{equation}
    \mathcal{L}_{\text{cell-cov}} = \frac{1}{N_c} \sum_{j=1}^{N_c} \text{ReLU} \left(1 - \sum_{i=1}^{N_s} p^{\text{cell}}_{ij} \cdot v^{\text{cell}}_{ij} \right)\label{eq:cell-coverage-loss}
\end{equation}

\textbf{Capacity Constraint Losses}
To encourage load balancing, we introduce a set of losses which penalize assigning excessive demand to a particular satellite or GW.
Let the load assigned to satellite $i$ be:
\begin{equation}
    L_i^{\text{sat}} = \sum_{j=1}^{N_c} p^{\text{cell}}_{ij} \cdot v^{\text{cell}}_{ij} \cdot d_j
\end{equation}

The gateway and satellite capacity losses are then defined as:
\begin{align}
    \mathcal{L}_{\text{capacity}} &= \frac{1}{N_s} \sum_{i=1}^{N_s} \text{ReLU} \left( \sum_{j=1}^{N_c} L_i^{\text{sat}} - C_{\text{max}} \right)\\
    \mathcal{L}_{\text{gw-cap}} &= \frac{1}{N_g} \sum_{k=1}^{N_g} \text{ReLU} \left( \sum_{i=1}^{N_s} p^{\text{gw}}_{ik} \cdot v^{\text{gw}}_{ik} \cdot L_i^{\text{sat}} - G_{\text{max}} \right)
\end{align}
As with the coverage loss, the ReLU zeros out the loss if our constraint is met.
Note that the coverage losses and load balancing losses work in opposition: we could guarantee coverage by covering all cells with all visible satellites, but that would result in too much load assigned to each satellite.

\textbf{Unique Assignment Losses}
These losses ensure that each satellite selects a single gateway and each ground cell is assigned to one satellite, producing unambiguous decisions suitable for real deployments.

\noindent\textit{Satellite-Gateway Assignment Exclusivity Loss  }
To simplify deployment, each satellite typically uses a single gateway within each snapshot. This loss penalizes fractional assignments across multiple gateways, encouraging clear, one-to-one connections for practical routing and scheduling.
\begin{equation}
    \mathcal{L}_{\text{sat-assign}} = \frac{1}{N_s} \sum_{i=1}^{N_s} \left| \sum_{k=1}^{N_g} p^{\text{gw}}_{ik} \cdot v^{\text{gw}}_{ik} - 1 \right|
\end{equation}

\noindent\textit{Cell-Satellite Assignment Loss } 
Similarly, we assume each ground cell is served by only one satellite at a time.
This maximizes coverage and load balancing by allowing satellites to focus on smaller groups of cells and simplifies the access for users.

\begin{equation}
    \mathcal{L}_{\text{cell-sat}} = \frac{1}{N_c} \sum_{j=1}^{N_c} \text{CrossEntropy} \left( p^{\text{cell}}_j,\; \arg \max_i p^{\text{cell}}_{ij} \right)
\end{equation}

\textbf{Total Loss Objective}  
The overall loss is a weighted sum of all individual constraint terms. The weights are tuned during training and selected based on performance and stability:
$\mathcal{L}_{\text{total}} = \sum_{i} \gamma_i \mathcal{L}_i$
The loss weights encode operator trade-offs between coverage and utilization and can be tuned to match service objectives.

Note that the loss used to train \system is unsupervised in the sense that there are no ground truth labels.  This is for two reasons: 
The first is that it is difficult to generate a large dataset of ``optimal'' orchestration decisions given the time required to solve the multi-dimensional assignment problem.
The second is that it allows us to incorporate the constraints directly into the losses.
Training ML models with constraints is known to be a difficult challenge~\cite{cotter-optimizationnondifferentiableconstraints-2019}.  Incorporating them into a weighted loss function is a useful method in this case.  We should note that the output of \system may still violate some of the constraints such as the capacity constraint.  This may be inevitable given the ratio of the total demand to the capacity of the system, but in this case the system applies a demand satisfaction ratio as described in Section~\ref{sec:motivation}.

\begin{figure}[t]
    \centering
    \includegraphics[width=\linewidth]{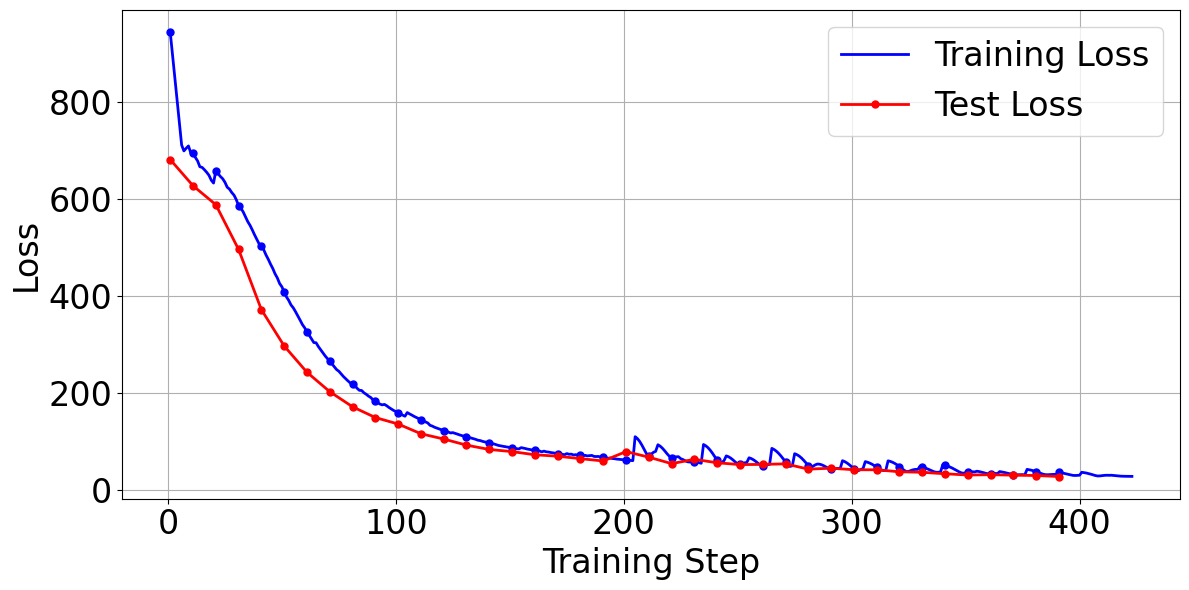}
    \caption{Total training and testing loss across time-varying graph snapshots.}
    \label{fig:loss_curve}
\end{figure}
\subsubsection{Training Across Time-Varying Graph Snapshots}
To reflect the dynamic nature of LEO satellite networks, the model is trained using a series of graph snapshots collected over multiple orbital cycles. 
Each snapshot captures a unique network configuration caused by satellite movement and evolving visibility to ground cells and gateways. 
We train on a schedule in two parts.  In the first part we train on consecutive snapshots, capturing temporal continuity.  Then in the second part, we introduce variability by training on random, non-continuous samples from a different part of the day.
We find that training with a batch size of one is most effective.
Each epoch consists of multiple training steps on a single snapshot.
Figure~\ref{fig:loss_curve} shows the loss curve on a small dataset of 40 training samples (20 each of continuous and random) with 20 training steps per epoch. The random phase begins at step 200.
The steady decline in loss reflects the \system's growing ability to generate valid, load-balanced assignments that satisfy coverage and capacity constraints.
During the continuous phase, we have a smooth loss curve due to using consecutive snapshots; this is because not much changes between training steps.  When we get to the random phase, each new sample results in a small increase in the loss function before adapting.  This is expected as these samples represent potentially different satellite configurations and demand patterns; however, these spikes are minor, decreasing, and do not alter the learning trend in the curve.

\subsection{Deployment Considerations}
\label{sec:deployment}

The trained GNN orchestration model can be deployed in the following configurations. They highlight some level of autonomy and operational control in LEO constellations.

  \noindent\textbf{Ground-Based Centralized Inference} 
    A centralized controller on the ground computes satellite-to-cell and satellite-to-gateway assignments using the full network state per second. It may transmit the updated decisions at a configurable frequency (10-20 seconds). This setup aligns with bent-pipe architectures and requires no onboard processing, enabling globally optimized coordination. 

 \noindent\textbf{Partial Onboard Inference } 
   To extend the autonomy and reduce the dependency on ground based controllers, the next step is to enable satellites to execute a cached version of the GNN model locally. Each satellite recomputes its assignments using local visibility and recent states, allowing it to adapt during ground link loss, gateway overload, or partial network failure. This model does not require full decentralization.

   A central question still exists: when do we push the model to satellites, and how often does that need to happen?  Typically, this is a one-time deployment, with infrequent updates unless the network undergoes major structural changes. As shown in Section~\ref{sec:evaluation}, minor demand shifts or localized failures can be handled through local adaptation without retraining.

   However, issues like decreased coverage, rising assignment failures, or divergence between predicted and actual load distributions may prompt fallback to onboard models or updated deployments from the ground. Operators usually make this decision based on real-time telemetry.
    Partial onboard inference enables a hybrid coordination model: leveraging centralized updates when possible, and relying on local inference during outages. In the future, ISLs may enable lightweight inter-satellite coordination, improving resilience under failures.

%% file: 4_evaluation.tex
\section{Evaluation}
\label{sec:evaluation}

\subsection{Experimental Setup}
\label{subsec:experimental-setup}

All simulations are conducted over some or all of the continental United States using realistic deployment parameters.
User demands are not simulated individually, but are instead aggregated into cells. Cells are defined by Uber’s H3 geospatial indexing system. The system uses 4,569 hexagonal ground cells, each covering approximately 26~km\textsuperscript{2} on average.
Each ground cell generates average of 20 Mbps, and for training, each satellite is assigned a maximum service capacity of 3 Gbps. A uniform demand model isolates orchestration effects from traffic variability, ensuring that load imbalance and coverage gaps reflect coordination behavior. The framework is agnostic to the demand model, with demand used only as a node feature.
The 54 gateway locations are based on crowd-sourced reports of real Starlink gateway locations~\cite{-starlinkgroundstation-2023}. The simulated satellite constellation is based on Starlink Phase I, consisting of 1,584 satellites arranged across 72 orbital planes, each containing 22 satellites.  Although the full constellation is modeled, only the subset of satellites visible over United States at any given timestep, typically around 120, are active satellites for North America. These orbits are inclined at $53^\circ$ and operate at an altitude of 550~km, yielding an average orbital period of 95 minutes.
Visibility between satellites and ground entities (cells or gateways) is determined by geometric line-of-sight constraints and a minimum elevation angle threshold of $25^\circ$, reflecting practical antenna steering and propagation constraints.
The training dataset consists of 24~hours of 20~second snapshots.  Of these 4320 snapshots, a disjoint subset is used for training and testing.  Both training and test sets consist of 50\% continuous samples and 50\% random samples.
All results presented (\ref{sec:evaluation}), for Optimizer, Local heuristics, Network based models and GNN variants, are obtained from the same test set to ensure consistency and comparability across the models.
Though \system works for general LSN deployments, for the purpose of evaluation, we only consider bent-pipe connections.  This means we only consider satellites which can see a GW.

\textbf{Model Variants and Baselines:}
We evaluate the proposed heterogeneous GNN-based coordination model, referred to as \system, against four representative baselines:

\noindent \textbf{L-Prox (Proximity Heuristic):} Each ground cell is assigned to the geographically closest visible satellite, without considering capacity constraints.  Likewise, all satellites connect to their closes visible GW.

\noindent \textbf{L-Demand (Demand-Greedy):}  Each ground cell is processed in order of highest demand, and assigned to the nearest satellite with available capacity. Satellites are scored using a distance-to-demand ratio, adjusted by current load. GW assignment is the same as L-Prox.

\noindent\textbf{Supervised ML (ML):} A static learning model trained on visibility and geographic features.

\noindent \textbf{Homogeneous GNN (HGNN):} A graph-based model treating all nodes identically, without role-aware reasoning~\cite{chen-deepreinforcementlearningbased-2025}.

\noindent\textbf{Centralized Optimizer (Opt):} A global assignment solver with full access to network state and constraints, used as an idealized performance upper bound.

\subsection{Performance: \system vs. Optimized Orchestration}
\label{subsec:tradeoff_gnn_opt}

This section compares the performance of \system against a centralized optimizer (Opt) that solves the satellite-to-cell and satellite-to-gateway assignment problem as a global combinatorial optimization described in Section~\ref{sec:motivation}.

\begin{figure}
    \centering
    \begin{minipage}{0.49\linewidth}
        \centering
        \includegraphics[width=\linewidth]{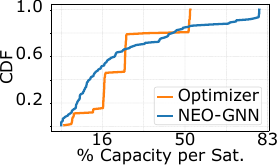}            
        \caption{\%$C$ per Sat. CDF}
        \label{fig:cdf_optimizer_gnn}
    \end{minipage}
    \hfill
    \begin{minipage}{0.49\linewidth}
        \centering
        \includegraphics[width=\linewidth]{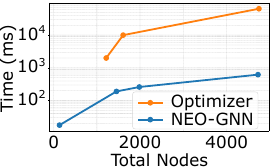}
        \caption{Runtime vs. Size}
        \label{fig:runtime_comparison}
    \end{minipage}
\end{figure}

Figure~\ref{fig:cdf_optimizer_gnn} shows a CDF of the percentage of satellite capacity used following the orchestration of Opt and \system.
Both systems are able to limit overloading of satellite nodes through proper load balancing, though \system has a bit more variance.
As shown in the next section, local and naive ML baseline fail to balance load effectively, resulting in demand satisfaction dropping. Note that both \system and Opt guarantee full coverage for the default scenario, where there are sufficient satellites.
Because they achieve 100\% coverage and all satellites loads are less than the capacity, we also achieve 100\% demand satisfaction.

Where \system greatly outperforms Opt is during deployment.
Figure~\ref{fig:runtime_comparison} shows the run time for different sizes of of deployments in terms of the total number of nodes (see Table~\ref{tab:quasi-global-runtime} for specific node breakdowns; note that the y-axis is in log scale). 
While the optimization solve time quickly grows to many 10s of seconds, \system's inference time remains on the order of milliseconds.
This is critical because the orchestration decisions must then be broadcast to the satellites for execution and operation.

Despite slightly higher load variance, \system matches the Opt on coverage and utilization while delivering orders-of-magnitude faster inference, enabling rapid reconfiguration in dynamic LEO environments.

\begin{figure*}
    \centering
    \begin{minipage}{0.3\textwidth}
        \centering
        \includegraphics[width=\linewidth]{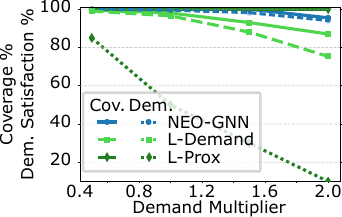}
        \caption{Coverage (Cov) and Demand Satisfaction (Dem.) vs. demand level}
        \label{fig:demand_scaling_dual_axis}
    \end{minipage}
    \hfill
    \begin{minipage}{0.3\textwidth}
        \centering
        \includegraphics[width=\linewidth]{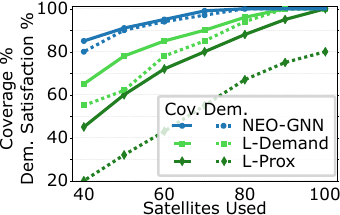}
        \caption{Coverage (Cov) and Demand Satisfaction (Dem.) \% vs. \# Sats.}
        \label{fig:satellite_scaling_breakdown}
    \end{minipage}
    \hfill
    \begin{minipage}{0.3\textwidth}
        \centering
        \includegraphics[width=\linewidth]{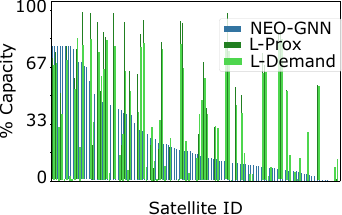}
        \caption{Per-satellite total demand}
        \label{fig:demand_dist_local}   
    \end{minipage}
\end{figure*}
\subsection{Baseline Comparison: \system vs. Local Heuristics}
\label{subsec:local_vs_gnn}

This section evaluates \system against two local heuristics: L-Prox (nearest-assignment) and L-Demand (greedy demand selection). These models are computationally simple, but lack network coordination and structural awareness.

\subsubsection*{\underline{Demand Scaling Evaluation}}
Figure~\ref{fig:demand_scaling_dual_axis} shows system behavior as total demand increases from $0.6\times$ to $2.0\times$ over the training baseline demand. \system consistently maintains over 99\% coverage and 95\% demand satisfaction across all levels. In contrast, while L-Prox maintains 100\% coverage (since the orchestration is demand independent) it and L-Demand degrade sharply beyond the $1.2\times$ demand threshold, with coverage gaps and unsatisfied demand increasing rapidly. Unlike the abrupt drops observed in the local heuristics, \system demonstrates smooth, predictable degradation under overload, highlighting its robustness.
This confirms that the structure-aware coordination logic in \system enables better resource sharing and avoids premature load saturation. In contrast, local decisions in L-Demand and L-Prox, made without awareness of broader network conditions, leads to imbalanced load distribution and coverage degradation.

\subsubsection*{\underline{Satellite Scaling Evaluation}}

We assess this by selecting subsets of satellites from the full constellation and treating only those as active during training and evaluation.
For each subset size, training is done using graph snapshots containing only the corresponding satellites, along with visible ground cells and gateways.
Figure~\ref{fig:satellite_scaling_breakdown} shows the resulting coverage and demand satisfaction for \system, L-Demand, and L-Prox.

As satellite count increases, \system consistently delivers superior performance across both coverage  and demand satisfaction.
\system reaches near-100\% coverage starting around 50 satellites and maintains it across all larger configurations. L-Demand and L-Prox trail behind, with coverage gaps persisting even at 80+ satellites.
\system also satisfies a notably higher fraction of the overall traffic demand
\system continues to scale effectively, maximizing utilization.

\subsubsection*{\underline{Load Distribution Comparison}}
To understand why, we look at the satellite distribution for a single snapshot.
Figure~\ref{fig:demand_dist_local} shows the total assigned demand per satellite under the three models. All models serve comparable total demand, however, \system distributes it significantly more evenly across available satellites. In contrast, both L-Demand and L-Prox exhibit a higher variance, with many satellites either overloaded or left idle due to uncoordinated, local decisions. This balanced load distribution explains \system's superior performance in earlier evaluations in demand and satellite scaling where it sustains higher coverage and demand satisfaction by efficiently utilizing all visible satellite capacity.
\system's structure-aware inference enables efficient assignment by incorporating local topology and peer activity through message passing, which balances demand across the constellation, prevents local congestion, and maintains full coverage under constraints. These results confirm that \system generalizes well to expanded constellations and effectively leverages additional satellite capacity, outperforming myopic heuristics 
through its scalable and adaptive coordination strategy.

\subsection{Network-Based Model Comparison: ML, HGNN}
\label{subsec:model_comparison_gnn}

\begin{figure*}
    \begin{minipage}{0.3\textwidth}
        \centering
        \includegraphics[width=\linewidth]{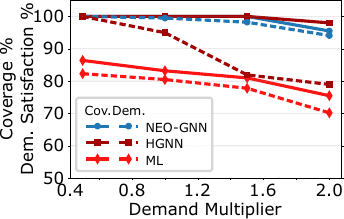}
        \caption{Coverage (Cov) and Satisfaction (Dem.) \% vs. demand level}
        \label{fig:demand_scaling_model_comparison}
    \end{minipage}
    \hfill
    \begin{minipage}{0.3\textwidth}
        \centering
        \includegraphics[width=\linewidth]{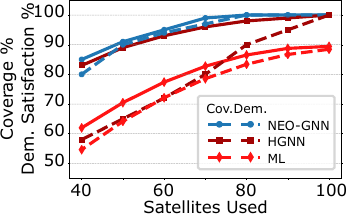}
        \caption{Coverage (Cov) and Satisfaction (Dem.) \% vs. \# Sats.}
        \label{fig:satellite_scaling_model_comparison_network}
    \end{minipage}
    \hfill
    \begin{minipage}{0.3\textwidth}
        \centering
        \includegraphics[width=\linewidth]{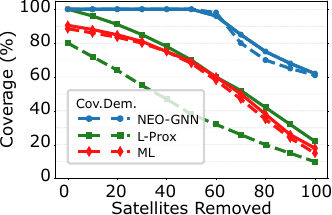}
        \caption{Increasing satellite failures}
        \label{fig:demand_vs_removed}
    \end{minipage}
\end{figure*}

This section evaluates \system against two network-based learning approaches: ML (supervised learning) and HGNN (homogeneous GNN). While both the models incorporate learned inference and consider a global picture, they lack role-specific reasoning and exhibit limited adaptability under dynamic network conditions.

\subsubsection*{\underline{Demand Scaling Evaluation}}
Figure~\ref{fig:demand_scaling_model_comparison} shows how model performance evolves as traffic demand scales from $0.5\times$ to $2.0\times$ the training baseline demand. As with the local heuristics, these network models also typically degrade in performance with increasing demand, despite being aware of global information.
The supervised learning model, ML, exhibits the worst performance in both coverage and demand satisfaction, illustrating the difficulty in using labels to train such a complex optimization problem.
On the other hand, HGNN has as good or slightly better coverage than \system as the demand increases, but it comes at the cost of lower demand satisfaction which falls to less than 80\% at $2\times$ the total load.
These results highlight \system's superior generalization and robustness under traffic load. Unlike HGNN and ML, which fail to enforce constraints under stress, \system maintains performance by balancing load and adhering to satellite visibility and capacity limits.

\subsubsection*{\underline{Satellite Scaling Evaluation}}
Figure~\ref{fig:satellite_scaling_model_comparison_network} shows how each model performs as satellite count increases.
With respect to coverage, \system reaches near-100\% coverage around 50 satellites and sustains it thereafter. HGNN improves to ~90\% coerage but saturates, while ML stays below ~85\% coverage regardless of satellite count.
With respect to demand, \system maintains over 95\% satisfaction throughout. While HGNN peaks near 85\% and  flattens, ML remains under 75\% satisfaction, unable to adjust to capacity growth.
Thus, \system reaches optimal coverage and satisfaction with as few as 50 satellites and maintains high performance across the full range. In contrast, while HGNN shows moderate improvement but plateaus early,  ML fails to capitalize on increased capacity due to its static inference logic.

\subsection{Resilience to Infrastructure Adaptation}
 \system leverages GNN features to remain resilient to changes in graph topology. Since the graph $\mathcal{G}_t$ is an \textit{input} during inference time, we can handle updates to the satellite constellation or ground segment deployment \textit{after} we have trained \system.
Figure~\ref{fig:demand_vs_removed} shows that \system continues to perform well even after up to 50\% of the satellites \textit{failure} (e.g. for satellites going offline).
In contrast, ML and L-Prox degrade rapidly, failing to maintain coverage or satisfy demand under failure. The degradation in \system remains smooth, without abrupt collapse, indicating effective load redistribution. This highlights its ability to maintain service continuity under dynamic conditions.
Specifically, \system sustains over 90\% coverage even with 40\% of satellite lost. ML and L-Prox drop rapidly, falling below 60\% at 50\% removal.
\system also retains over 80\% satisfaction after 50\% satellite failure. ML and L-Prox drop to 50\% or less, failing to recover under loss.
\system is robust because of its ability to redistribute demand efficiently among the remaining satellites while respecting system constraints.

\system also exhibits strong scalability under infrastructure expansion. Even when the active satellite pool increases by 25\% during deployment (not considered in training), the model maintains consistent coverage, balanced assignments, and adherence to system constraints,without retraining or reconfiguration. This shows that the learned coordination policy scales to larger constellations and handles added capacity  without instability or oversubscription.
Overall, the results show that \system is resilient to partial infrastructure failures and scalable to future growth, offering fault-tolerant and forward-compatible orchestration  in dynamic LEO networks.

%% file: 5_related_work.tex
\section{Related Work}
\label{sec:related_work}

LEO-based NTNs are a cornerstone of 5G/6G evolution, particularly in regions lacking terrestrial infrastructure \cite{liu-leosatelliteconstellations-2021}. 
Much of LSN satellite research has dealt with load-balancing and routingin \textit{within} the satellite constellation~\cite{han-loadbalancingroutingleo-2023,liu-loadbalancingrouting-2021,zhi-loadbalancingrouting-2025}.
These works do not consider the end-to-end problem; the traffic matrix already exists.  In this work we handle both the ingress and egress part of the problem as well.
Machine learning methods have been explored for routing and load balancing, but they generally struggle to enforce constraints and ensure full coverage under dynamic conditions \cite{zhao-demandawarebeamhopping-2024,ding-fastconvergencereinforcementlearning-2023}.

Graph Neural Networks (GNNs) have been applied to satellite routing, scheduling, and handover management~\cite{gao-topologycompresseddatadelivery-2025,liu-routingsmallsatellite-2021,casadesus-vila-autonomouscooperationheterogeneous-2024,lee-handoverstrategyleo-2025}, but most assume homogeneous node behavior, limiting their ability to capture role-specific constraints in mixed-role systems. 
Heterogeneous GNNs (HetGNNs) are designed to model graphs with multiple node and edge types, and they have demonstrated success across domains such as recommendation understanding~\cite{shi-heterogeneousgraphneural-2022,zhang-heterogeneousgraphneural-2019} and region-level geospatial embedding \cite{zou-learninggeospatialregion-2024}. Nevertheless, their potential in LEO resource coordination—where distinct roles define system constraints—remains largely untapped.
Alternative methods such as Ant Colony Optimization (ACO) \cite{zhi-loadbalancingrouting-2025}, Federated Learning \cite{razmi-onboardfederatedlearning-2021}, demand-aware beam hopping and power control \cite{zhao-demandawarebeamhopping-2024}, and segment routing have been proposed for satellite networks. Each method offers valuable insights, they struggle to simultaneously address the utilization, coverage, and real-time constraints central to LEO network orchestration.

%% file: 6_conclusion.tex
\section{Conclusion}
\label{sec:conclusion}

This paper introduces a novel heterogeneous Graph Neural Network (GNN) based learning framework, \system for LEO network orchestration. 
By modeling the network as a dynamic, typed graph, our GNN captures the distinct roles of satellites, ground cells, and gateways. This enables type-specific message passing and constraint-aware inference that allows \system to efficiently balance traffic demand satisfaction and cell coverage simultaneously, a key feature that is hard to deliver by existing schemes.
Our evaluation demonstrates that \system achieves near-optimal performance, ensuring full coverage, balanced load distribution to yield high network utilization, while offering significant computational advantages over centralized optimization, make it suitable for practical, real-time deployments.